\documentclass[]{emulateapj}
\shorttitle{Imbalanced Turbulence}
\begin{document}
\title{Strong Imbalanced Turbulence}
\author{A. Beresnyak, A. Lazarian}
\affil{Dept. of Astronomy, Univ. of Wisconsin, Madison, WI 53706}
\email{andrey, lazarian@astro.wisc.edu}

\def\L{{\Lambda}}
\def\l{{\lambda}}

\begin{abstract} 
We consider stationary, forced, imbalanced, or cross-helical MHD Alfv\'enic turbulence where
the waves traveling in one direction have higher amplitudes than the opposite waves.
This paper is dedicated to so-called strong turbulence, which cannot be
treated perturbatively. Our main result is that the anisotropy of the weak
waves is stronger than the anisotropy of a strong waves.
We propose that critical balance, which was originally conceived as
a causality argument, has to be amended by what we call a propagation
argument. This revised formulation of critical balance is able to handle
the imbalanced case and reduces to old formulation in the balanced case.
We also provide phenomenological model of energy cascading and discuss
possibility of self-similar solutions in a realistic setup of driven
turbulence.
\end{abstract}


\section{Introduction}
MHD turbulence appears in the dynamics of conductive fluid in a generalized settings
with large Reynolds numbers, or low physical dissipation.
It is ubiquitous in the interstellar and intracluster medium, Earth
magnetosphere, solar wind, accretion disks, etc. In fact, it is laminar flows that constitute
exception in astrophysics, while, generically, astrophysical fluids are
turbulent.

The study of MHD turbulence has been an old challenge. First
attempts to address it were classical papers by Iroshnikov (1963) and
Kraichnan (1965) (henceforth IK model). A good account for the state
of the field could be found in
Biskamp (2003). Usually turbulence is subdivided into weak and
strong, depending on the strength of non-linear interaction. While weak
MHD turbulence allows analytical perturbative treatment
(Ng \& Bhattacharjee 1996, Galtier et al 2002, Chandran 2005),
the progress in understanding strong turbulence came primarily from
phenomenological and closure models which were tested by numerical simulations.

Important theoretical works on strong MHD turbulence include
Montgomery \& Turner (1981), Shebalin, Matthaeus, \& Montgomery (1983),
Higdon (1984). Those clarified the
anisotropic nature of the energy cascade and paved the way for further
advancement in the field.
The study by Goldreich \& Sridhar
(1995, henceforth GS95) identified the balance between perturbations
parallel and perpendicular to the local direction of magnetic field,
i.e. ``critical balance'', as the key component of dynamics in strong magnetic
turbulence.  Although it dealt with incompressible MHD turbulence, GS95
also influenced further studies of compressible turbulence
(e.g. Lithwick \& Goldreich 2003).  In particular, it identified the
dominant role of Alfv\'enic perturbations for cascading of slow modes,
which later confirmed with numerical simulations in both
weakly and strongly compressive media (Cho \& Lazarian 2002, 2003).

Being a mean-field model in the spirit of Kolmogorov (1941), GS95
predicts the velocity and magnetic field fluctuation strengths and their anisotropies,
in terms of the local dissipation rate. Even though recently the simple
dynamical model of GS95 came under criticism (see, e.g., Boldyrev 2006,
Gogoberidze 2007) with the motivation to explain the deviations from
GS95's -5/3 spectrum in numerical simulations (see, e.g., Muller, Biskamp \& Grappin 2003),
we feel that it does provide a good insight into MHD turbulence.
The model that we present in this paper, is similarly to Kolmogorov or GS95,
a {\it mean field model}, which does not account for any local dynamical
effects or intermittencies (cf Beresnyak \& Lazarian 2006). These effects are beyond the scope
of this paper and will be addressed elsewhere.

While balanced MHD turbulence enjoyed much attention, the opposite
regime, i.e., imbalanced turbulence, was less developed.  The
analytical results were obtained for weak imbalanced turbulence
(Galtier et al. 2002, Lithwick \& Goldreich 2003) though they are
applicable in a rather narrow range of imbalances.

So far, the simulations of strong imbalanced turbulence
were limited to rather idealized set ups (Maron \& Goldreich 2001, Cho, Lazarian \& Vishniac 2002),
which did not allow making definitive conclusions about its properties. In this paper, in order to
test our analytical model, we provide simulations which go a step further compared to the 
aforementioned studies. We hope that future higher resolution simulations will provide the definitive
test for alternative models of imbalanced turbulence. 

We think that the best experimental data on the imbalanced regime is currently available from observations
of solar wind turbulence (e.g., Horbury 1999). This data,
collected by spacecrafts, is consistent with Kolmogorov $-5/3$ spectrum,
but does not provide sufficient insight
into the anisotropy with respect to the local magnetic field.
The imbalanced turbulence is not a rare exception, on the contrary,
such processes as preferential decay of a weaker wave and the escape
from the regions that generate perturbations make the imbalanced
turbulence ubiquitous. It goes without saying that models
of such turbulence are much needed in astrophysics.

An attempt to construct the model of stationary {\it strong} imbalanced
turbulence was done in Lithwick et al. (2007), henceforth LGS07.
This model assumes the strong GS95-type cascading of both large and small amplitude oppositely
moving modes. In what follows, we propose a different model of imbalanced turbulence. In \S 2 we revisit
the critical balance argument and discuss a new process of cascading, which we relate to the
process of propagating of Alfv\'enic perturbations in the field wandering which is induced by
the oppositely moving wave. Using this "propagation cascading" in case of the
balanced turbulence we recover GS95 relations. However, for the pronounced imbalance between the
oppositely moving waves, this process results in a picture which is different from that in GS95.
In \S 3 we 
discuss the scaling relations for the turbulent cascade that follow from our model and in
\S 4 we compare our
predictions with our 3D numerical simulations of the imbalanced cascade. In \S 5 we compare our
results with other studies.

\section{Critical balance revised}

The ``perpendicular cascade'',
a concept which was rigorously developed in theory of weak Alfv\'enic
turbulence (Ng \& Bhattacharjee, 1996, Galtier et al, 2002), was a theory
of nonlinear interacting Alfv\'enic waves which, due to the particular
dispersion relation of the waves, conserved wave frequencies $\omega=k_\|v_A$.
It deeply contrasted with earlier Iroshnikov-Kraichnan models where, due to 
the assumed isotropization,
the parallel wavenumber will be of the order of the total wavenumber and
the frequency $\omega=k_\|v_A\approx kv_A$ changes with $k$.
The perpendicular cascade, however, makes the cascade {\it stronger}
and not weaker, while the energy goes downscale. This raised a question of what happens
when the perturbation theory breaks down and the turbulence becomes strong.
GS95 argued that the turbulence will stay on the edge of being strong, because of the uncertainty
relation between the cascading timescale and the wave frequency $t_{\rm cas}\omega\sim 1$.
Their EDQNM closure model contained explicit ad hoc term that allowed for the increase
of frequencies of the interacting wavepackets\footnote{In this paper we will use wavevectors
$k_\|$, $k_\perp$ and length scales $\L=1/k_\|$, $\l=1/k_\perp$ interchangeably. While
wavevector representation highlights geometry and conservation laws, the length scales
are measured quantities obtained from statistical averaging of numerical
or observational data. We will also use ``waves'', ``wavepackets'' and ``eddies'' interchangeably.}.
Since this process of increase of $k_{\|}$ comes essentially from the
irreversibility of energy cascading downscale, we call it the {\it causality effect}.
This way nonlinear interaction stays marginally strong by controlling
the anisotropy. But is this the only way to increase frequency or $k_{\|}$?
In this paper we advocate to supplement the causality effect with a different mechanism
which works when the cascading of the eddy is done by the counter-eddy, which is
tilted with respect to the mean field due to the different definition of the local
mean field at different scales. The details are following.

Indeed, the parallel wavenumber for an Alfv\'enic eddy in strong turbulence actually
depends on how the local mean field is defined. Maron \& Goldreich (2001)
have shown that the Alfv\'enic eddy propagates along the field lines that
are defined by the average field {\it plus} the field of the counter-wave.

Let us try to calculate the characteristic uncertainty in $k_\|$ that comes
from this effect in the {\it balanced} case. In general, it will depend on eddy's
polarization, but at maximum it will be around $k\sin \theta$, where $\theta$
is the angle of the field wandering, or $k_\perp \delta b(l)/v_A$ ($b$ is the magnetic
field perturbation in the Alfv\'enic units). What is the characteristic scale $l$
that enters this expression? We can argue that in the case of a {\it local balanced} turbulence
there is only one designated scale, the one we consider cascading from. Therefore,
in the case of balanced turbulence we expect $\delta k_\|=k_\perp\delta b(1/k_\perp)/v_A$.
This coincides with the well known critical balance, derived from causality (indeed, in GS95
the cascading rate is of the order of eddy turnover rate $k_\perp\delta v$) .
We were able to reproduce causal critical balance of strong MHD turbulence
(in the balanced case) by argumentation that involved field wandering.
We call this effect, that comes from field wandering, the {\it propagation}
critical balance and argue below that
it will provide qualitatively new picture in the imbalanced case. 

Let us now consider the imbalanced case. We will adhere to the same
``eddy'' ansatz as GS95 in that for every transverse scale of a wave
there is a characteristic longitudinal scale. This is the physical meaning behind terms
``eddies'' or ``wavepackets''. We will digress from GS95 in a natural way,
postulating that both waves have different anisotropies, i.e. the dependence
of longitudinal scale $\L$ to transverse scale $\l$ is different for each kind of wave.
This situation is presented in Fig. 1, where some arbitrary
longitudinal scale $\L^-$ corresponds to the {\it two different} transverse scales,
$\l_1$ for weak wave $w^-$ and $\l_2$ for strong wave $w^+$.
$\L^+$ is a longitudinal scale of $w^+$ wave having
transverse scale $\l_1$. Since we originally decided to consider strong turbulence,
let us assume that at least the $w^-$ is being strongly
cascaded by $w^+$. In this case the most effective mixing
of $w^-$ on scale $\l_1$ will be obtained through $w^+$ motions
that are on the same scale\footnote{This is the assumption that is often employed in a strong turbulence,
such as Kolmogorov model of incompressible hydro turbulence, or GS95.}.
The longitudinal scale for $w^-$ will be provided by causal critical balance, since
its cascading is fast.

\begin{figure}
\figurenum{1}
\plotone{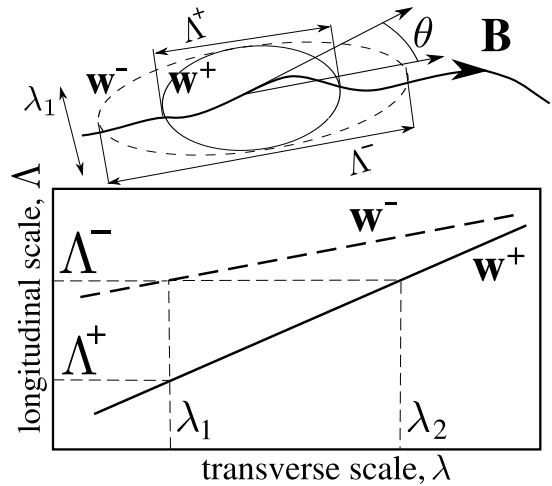}
\caption{Upper: a ${\bf w}^+$ wavepacket, produced
by cascading by ${\bf w}^-$ wavepacket is aligned with respect
to ${\bf w}^-$ wavepacket, but misaligned with respect
to the local mean field on scale $\lambda_1$, by the angle $\theta$.
Lower: the longitudinal scale $\L$ of the wavepackets,
as a function of their transverse scale, $\l$; $\L^+$, $\L^-$, $\l_1$, $\l_2$
are the notations used in this paper.}
\label{anis_cartoon}
\end{figure}

The cascading of $w^+$ is somewhat more complicated. Since
the amplitude of $w^-$ is not large enough
to provide strong perturbations in $w^+$, the $w^+$ will be perturbed weakly, and
the cascading timescale will be diminished according to the ``strength''
of the $w^-$, just like it does in weak turbulence. Moreover, now the $w^-$ eddies
will be cascading $w^+$ eddies with similar longitudinal scales,
which is the generic feature of weak cascading.

The perturbations provided by $w^-$ will have a transverse scale of $\l_1$.
In other words, the energy of $w^+$ will be transferred between $\l_2$ and $\l_1$.
What determines the longitudinal scale for {\it cascaded} $w^+$?
This is the central question of this paper. We argue, that this
longitudinal scale will be determined by the propagation critical balance,
in the following way.
The wavepackets of $w^-$ are strongly aligned to the mean field
on scale $\l_1$, therefore they are randomly oriented with respect
to the mean field at a larger scale $\l_2$.
The RMS angle of wavevector of $w^-$ eddies with respect to mean field on $\l_2$ will be around
$\theta\approx\delta b^+(\l_2)/v_A$. This slant of $w^-$ wavepackets will determine
the increase of $k_\|$ for newly cascaded $w^+$ packets at $\l_1$ (see Fig. 1).

Finally, one can verify that the increase of $k_\|$ of $w^+$ due to causality
is smaller (due to the fact that cascading is rather slow),
and can be neglected. Also the increase of $k_\|$ in $w^-$ eddies due to propagation effect
will be negligible due to very weak wandering provided by $w^-$ field.
In this picture we expect the $w^-$ wavepackets on scale $\l_1$ to closely follow
mean field lines on scale $\l_1$. Essentially, this will require $\L^->\L^+$.
Since the amplitude of the $w^+$ is much larger than that of the $w^-$,
it provides much stronger increase of $k_\|$ by propagation, effectively
making the $w^+$ eddy {\it less anisotropic} than $w^-$ eddy and
validating this assumption.

Summarizing the result of this section, the new interpretation
of critical balance in the strongly imbalanced case is that the
$k_\|$ of the weak wave increases due to the {\it finite lifetime}
of the wave packet, while in the strong wave it increases
due to the field wandering of the strong wave itself
{\it on larger scales}. This effect does not contradict the exact
MHD solution of the wave propagating in one direction, because
it requires the oppositely propagating wave as an intermediary.
The de-alignment of cascaded strong wave is possible because
the weak wave, acting as cascading agent, is strongly aligned
with the field lines on scale which is {\it different} (smaller)
than the scale of the strong wave it is acting upon.

\section{Phenomenology}

We proceed with phenomenological treatment of imbalanced turbulence
using energy cascading rules for each wave and the new rules for
anisotropies developed in previous section. We denote Els\"asser wave
amplitudes in a sense of RMS values as $w^+$, $w^-$, where
``$+$'' corresponds to the strong wave and ``$-$'' to the weak wave.
Longitudinal and transverse scales are $\Lambda$ and $\lambda$ correspondingly,
and $\epsilon^\pm$ are energy fluxes for both waves.

Since we have different anisotropies for both waves, we are dealing with
4 quantities which are $w^+(\l)$, $w^-(\l)$, and $\Lambda^+(\l)$, $\Lambda^-(\l)$,
so we will need 4 equations to solve for
these functions. There are two equations for the constant
flow of energy in $k$-space and other two equations that determine frequency
uncertainties, or $\Lambda$s.
 
As was explained in \S 2, the longitudinal scale for weak wave
will be defined by causality, or,

$$\Lambda^-(\l_1)=v_A\left(\frac{w^+(\lambda_1)}{\lambda_1}\right)^{-1}. \eqno{(1)}$$

On the other hand, the strong wave's $\L$ will be defined by propagation,
as

$$\left(\frac{\L^+(\l^*)}{\l_1}\right)^{-1}=\frac{w^+(\l_2)}{v_A} \eqno{(2)}$$

(while \S 2 arguments assume $\l^*=\l_1$, it could actually be between $\l_1$ and $\l_2$,
see discussion below).
The weak wave energy flux will be determined by the strong shearing
of the strong wave as

$$\epsilon^-=\frac{(w^-(\l_1))^2 w^+(\l_1)}{\l_1}, \eqno{(3)}$$

while for the strong wave the energy flux will be decreased by the
wave perturbation strength as

$$\epsilon^+=\frac{(w^+(\l_2))^2 w^-(\l_1)}{\l_1}\cdot\frac{w^-(\l_1) \L^-(\l_1)}{v_A\l_1}\cdot f(\l_1/\l_2). \eqno{(4)}$$

Here $f(\l_1/\l_2)$ is a factor which will account for the fact that our weak cascading
is somewhat nonlocal and could also include the dependence on spectral slopes
(see Galtier et al 2002), though, for the sake of simplicity, in this section
we will assume $f=1$.

Let us now discuss a self-similar solutions of (1-4) which we seek in the form of

$$w^\pm(\l)=w^\pm_0(\l/L)^{\alpha_\pm}, \mbox{\ \ \ } \L^\pm(\l)=\L^\pm_0(\l/L)^{\beta_\pm}. \eqno{(5)}$$

The scaling exponents are calculated as $\beta_+=\beta_-=\beta$, $\alpha_+=1-\beta$,
$\alpha_-=(1-\alpha_+)/2$, here $\beta$ is arbitrary.
The correspondence with GS95 is obtained with $\beta=2/3$.

Let  us discuss the applicability of these solutions to the
realistic imbalanced turbulence. Quite surprisingly,
the self-similar solution
can not be applied to the imbalanced turbulence driven
isotropically on the outer scale. This will
require $\L^+_0(L)=\L^-_0(L)$ which is not possible, because equations
(3), (4) and (1) will give $\epsilon^+/\epsilon^-=1$.
This situation is exacerbated by the fact, that equations (1-4)
are nonlinear, so it is not possible to combine several
known solutions to satisfy boundary conditions.

Suppose, however, that self-similar solutions are
realizable with {\it some} outer scale boundary
conditions.
The way to uphold correspondence with GS95
$\beta=2/3$ is to take critical balance (1) and (2)
at outer scale for weak and strong wave respectively
and modify (2) such as both conditions are in agreement.
This is achieved, e.g., by replacing $\Lambda^+$
with geometrical average of $\Lambda^+$ and $\Lambda^-$.
Note that $\beta=2/3$ is also the one that has corresponding
spectral slopes $\alpha^\pm$ equal to each other and
a Kolmogorov value of $1/3$. If $\beta<2/3$ the spectra
are meeting and the self-similarity is broken.

With the solution, described above, we have
$\Lambda^-_0/\Lambda^+_0=\epsilon^+/\epsilon^-$
and $\lambda_2/\lambda_1=(\epsilon^+/\epsilon^-)^{3/2}$. Note that this
solution is realized not necessarily due to a different anisotropy of
$w^+$ and $w^-$ driving, but in a quite more general setting where
$\L^-_0(L)$ and $\L^+_0(L)$ play the role of
parameters of the asymptotic power-law solution. Indeed,
if we drive turbulence with the same anisotropy on outer scale,
there will be a non-power-law section of the solution that will
ensure that $\l_2/\l_1>1$, to fulfill $\epsilon^+/\epsilon^->1$,
which, given enough inertial range, will transit into the
asymptotic power-law solution. We believe that this situation
is realized, e.g., in the solar wind, where the counter-waves
are generated by reflection on density inhomogeneities on large
scales and thus have the same outer scale anisotropy as the direct waves.

\section{Factors,  relaxation to the steady state, viscous scale and limiting cases}

The equation (4) is written, assuming that cascading is weakened
in a way similar to how it is weakened in the weak Alfv\'enic turbulence. This
assertion does not come from any analytical argument, since
our turbulence could not be treated perturbatively. Due to the
fact that we have two length scales, $\lambda_1$ and $\lambda_2$
and two amplitudes, $w^+$ and $w^-$ there are several dimensionally
correct ways to write dissipation proportional to the fourth order 
of amplitude. We stuck with expression (4), as it seemed most
plausible. On the second note, the cascading in weak turbulence
is nonlocal in a sense, that the dissipation rate depends not
only on the local values of perturbation, but on the slope as
well (Galtier et al 2002, Lithwick \& Goldreich 2003).
Furthermore, in our model, the cascading of $w^+$ is somewhat
nonlocal, if $\lambda_1\neq\lambda_2$. The energy is being
transferred not in the local vicinity of one particular
scale, but between two different scales. In principle, this
would require a logarithmic correction factor such as
$f=1+f_{0}\log(\lambda_2/\lambda_1)$.

Strong local driven turbulence have steady-state relaxation timescales which
are usually equal to the dissipation timescale. Not only the $w^+$ cascading
time is much larger (see (4)),
but there is some nonlocality and slope dependence that,
in principle, could carry information upscale and make
the establishment of the steady state even longer.
With nonlinear dissipation timescale being very large, the
linear damping or viscosity will become important earlier, diminishing
inertial range. These factors make numerical study
of imbalanced turbulence extremely challenging.
Also, in a realistic compressible turbulence the parametric
instability or steepening become relatively more important,
as the turbulent cascading rates decrease.

Let us now discuss the onset of viscosity at small
scales. If our turbulence would be completely local
both in sense that same scales of ``$+$'' and ``$-$'' waves
cascade each other and in a Kolmogorov sense that energy
is transferred locally, that would mean equal viscous
scale for both waves and that the ratio of amplitudes must satisfy
$w^+/w^-=(\epsilon^+/\epsilon^-)^{1/2}$. Our model, however,
is somewhat nonlocal, and, according to our picture of cascading,
presented on Fig. 1, strong waves $w^+$ on scale $\l_2$ are cascaded
by weak waves $w^-$ on a smaller scale $\l_1$. Once the viscous dissipation
becomes important for $w^-$ on $\l_1$ scale, the amount of $w^+$ waves, cascaded
from $\l_2$ to $\l_1$ will strongly decrease. Therefore, the $w^+$ will have
shorter inertial interval than $w^-$. This effect is observed on numerical
spectra on Fig. 2. Since the viscous scales are different for $w^+$ and $w^-$,
the amplitude ratio mentioned above will be offset by some power of $\l_1/\l_2$.
Designating viscous scale as $\l^\pm_{\nu}$
and assuming $\l^+_{\nu}=\l^-_{\nu}(\l_2 / \l_1)^{1/2}$
we obtain $w^+(\l^+_\nu)/w^-(\l^-_\nu)=(\epsilon^+/\epsilon^-)^{1/2}(\l_2/\l_1)^{1/2}$.
If we use the self-similar, GS95-like solution from \S 3, i.e.
$\lambda_2/\lambda_1=(\epsilon^+/\epsilon^-)^{3/2}$, we get
$w^+(\lambda^+_\nu)/w^-(\lambda^-_\nu)=(\epsilon^+/\epsilon^-)^{1.25}$
(there is no guarantee, however, that a self-similar solution will
be established in a limited inertial range).

\begin{figure}
\figurenum{2}
\plotone{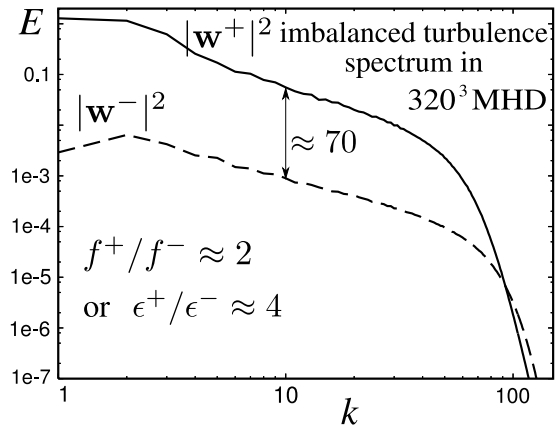}
\caption{The power spectrum of
energies for both wave species in an imbalanced forced
incompressive $320^3$ numerical simulation.
The energy input ratio $\epsilon^+/\epsilon^-$ was about 4.}
\label{spectrum}
\end{figure}

The limit of vanishing weak wave is understood as follows.
When $\epsilon^-$ goes to zero,
the steady-state relaxation time goes to infinity, in other words,
the steady state is never established and the amplitude of the
strong wave grows indefinitely from driving. Also, as $\l^+_{\nu}$
increases and approaches the outer scale, the viscous dissipation
becomes more important than the nonlinear dissipation and the cascading
stops, with the strong wave being present only on outer scale
and being dissipated viscously.

\begin{figure}
\figurenum{3}
\plotone{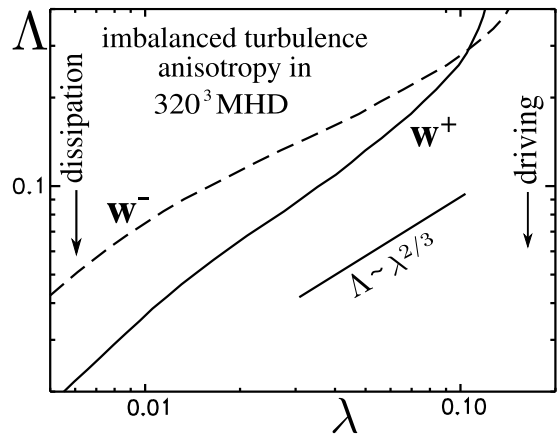}
\caption{The $\L(\l)$ dependence in the same simulation as in Fig.~2,
the length scales are in the units of the cube size.
The $\L\leftrightarrow\l$ correspondence is established by the
transverse and longitudinal second-order structure functions.}
\label{anisotropy}
\end{figure}

Finally, we would like to present our preliminary numerical results
of imbalanced driven turbulence. 
Both waves were driven independently with the same isotropy
on the outer scale and different amplitudes. The energy input
ratio $\epsilon^+/\epsilon^-$ was about 4. We evolved the simulation
for more than 30 Alfv\'enic times unless energies of both species,
their spectra and anisotropies has become stationary. The existence
of the stationary state of imbalanced forced turbulence is one of
the main assumptions in this paper as well as in LGS07.
The resulting spectra and anisotropies are shown in the Figs.~2 and 3.
The ratio of the total $|w^+|^2$ to $|w^-|^2$ ``energies''
was around a 100. In the middle of the inertial interval the ratio
of the power spectra was around 70 (LGS07 is predicting this
ratio as $(\epsilon^+/\epsilon^-)^2\approx 16$).  
The anisotropy curves (Fig. 3) diverge from outer scale,
as we suggested in \S 3. The $w^-$ perturbations exhibit
extreme anisotropy on small scales, while $w^+$ anisotropy is less
pronounced. The inertial range of $w^-$ is somewhat
larger than that of $w^+$ (Fig. 2), which is consistent with our model above. 
The effect of {\it pinning}, i.e.,
the equality $|w^+|^2=|w^-|^2$ at the dissipation scale,
predicted for the weak imbalanced turbulence by Lithwick \& Goldreich (2003)
is not observed in our simulations either\footnote{Our $\epsilon^+/\epsilon^-$ ratio, however,
is beyond the applicability of Lithwick \& Goldreich (2003).}.

At present, we think that different anisotropies of $w^+$ and $w^-$
perturbations, and, most notably, the {\it weaker}
anisotropy of $w^+$, is the best available evidence
favoring our model. Note that the weaker anisotropy of $w^+$ is hard
to explain in the framework of GS95 where parallel scale is equal
to the product of the cascading timescale and the Alfv\'en speed.

We will present more extensive set of imbalanced turbulence simulations
and a careful comparison with existing models in a future paper.

\section{Discussion}
The first treatment of imbalanced case of strong anisotropic turbulence,
LGS07, uses subtle argumentation in order to support the notion that anisotropies for
both waves will be the same. The main argument was that
even though the strain rate of the weak wave
will be smaller, it will be applied coherently, thus
resulting in large perturbations in the strong wave.
Therefore, as LGS07 argued, the old critical balance
will give the same longitudinal scale for both waves.
We feel that this argument is somewhat forced.
Indeed, the coherent application of the strain
does not necessarily mean cascading. While in case of the
strong shearing there's only one longitudinal scale,
in the case of weak shearing there are two:
the {\it correlation} scale and the {\it coherence} scale.
This is clearly seen in the case of weak Alfv\'enic turbulence
when the correlation scale is essentially determined by
the outer scale and is not changed by cascading, while the
coherence scale is determined by nonlinear cascading and is
typically much larger than correlation scale
(in fact this inequality is the condition of applicability
of weak turbulence). In weak turbulence the large coherence
scale does not lead to any unusual increase in cascading.
It is not at all obvious why this situation should change
in the case of strong turbulence, considered in LGS07.
Another shortcoming of LGS07 model is that, while being
completely local in both $k_\|$ and $k_\perp$, it dictates
different nonlinear cascading timescales for ``$+$'' and ``$-$'' waves,
which leads to inconsistency near the dissipation scale. 
Indeed, LGS07 has $w^+/w^-=\epsilon^+/\epsilon^-$, while
the local model must have $w^+/w^-=(\epsilon^+/\epsilon^-)^{1/2}$
to ensure transition into viscous/resistive dissipation (cf \S 4).

We note, that earlier closure models for isotropic MHD turbulence used phenomenological
``relaxation of triple correlations'' (see, e.g., Pouquet, Frisch,
Leorat, 1976) with Alfv\'enic timescale, which effectively lead to
the weakening of interaction and the IK $-3/2$ spectrum.
This approach introduced the idea that large scale field can lead to wave decorrelation
and the increase of wave frequency and $k_\|$. In this paper the increase of $k_\|$ is
obtained in a different way, namely, it is not dictated
by the full background field ${\bf B}$ and the Alfv\'enic timescale $(kv_A)^{-1}$, that is derived
from it, but by a perturbation $\delta {\bf B}$, which describes a difference between
local field averaged on two different scales, $\l_1$ and $\l_2$ (see Fig. 1). 

The influence of strong turbulence field wandering on the $k$
vector of the wave was proposed in Farmer \& Goldreich (2004) in
the context of the cascading of quasi-parallel modes, where
it was necessary to obtain non-zero $k_\perp$ in order the wave
to dissipate (see also Beresnyak \& Lazarian, 2008).
In our problem the situation is, in a sense, reversed,
as we need to obtain the increase of $k_\|$. 
In both cases the decorrelation seems to be a feature of strong
turbulence that does not directly appear in perturbative
calculations.

\acknowledgments
AB thanks IceCube project for support of his research.
AL acknowledges the  NSF grant AST-0307869 and the support from
the Center for Magnetic Self-Organization in Laboratory and Astrophysical
Plasma.

\end{document}